\documentclass{aa}
\usepackage{graphics}

\begin{document}

   \thesaurus{23         
              (04.01.1)} 
   \title{Search and Discovery Tools for
        Astronomical On-line Resources and Services}

    \titlerunning{Search and Discovery Tools}


\author{
    Daniel Egret\inst{1}
\and
    Robert J. Hanisch\inst{2}
\and
    Fionn Murtagh\inst{1,3}
          }

   \mail{Daniel.Egret@astro.u-strasbg.fr}

   \institute{CDS, Observatoire astronomique de Strasbourg, UMR 7550,
11 rue de l'Universit\'e, F-67000 Strasbourg, France
\and
  Space Telescope Science Institute, 
  3700 San Martin Drive, Baltimore, MD 21218, USA
\and
  School of Computer Science,
  The Queen's University of Belfast, 
  Belfast BT7 1NN, Northern Ireland
}

   \date{Received 5 January 2000 / \today}

\maketitle

\begin{abstract}

A growing number of astronomical resources and data or information
services are made available through the Internet. 
However valuable information is frequently hidden
in a deluge of non-pertinent or non up-to-date documents.
At a first level, compilations of 
astronomical resources provide
help for selecting relevant sites.
Combining yellow-page 
services and meta-databases of active pointers may be
an efficient solution to the data retrieval problem.
Responses generated 
by submission of queries
to a set of heterogeneous resources are
difficult to merge or cross-match, because
different data providers generally
use different data formats: new endeavors are under way
to tackle this problem.
We review the technical challenges involved
in trying to provide general search and discovery
tools, and to integrate them through upper
level interfaces.


\end{abstract}

%

\section{Introduction}

How to help the users find their way through the jungle of 
information services is a question which has been raised since the 
early development of the WWW (see e.g., Egret \cite{jungle}), 
when it became clear that a
big  centralized system was not the efficient way to go. 

Obviously the World Wide Web is a very powerful medium 
for the development of distributed resources: 
on the one hand the WWW provides a 
common medium for all information providers -- the language 
is flexible enough 
so that it does not bring unbearable constraints on existing 
databases -- on the other hand the distributed hypertextual 
approach opens the way to navigation and links
between services (provided a  minimum of coordination can be achieved). 
Let us note that it has been already widely demonstrated that 
coordinating spirit is not out of reach in a small community such 
as astronomy, largely sheltered from 
commercial influence.

Searching for a resource (either already visited, or 
unknown but  expected),
or browsing lists of existing services 
in order to discover
new tools of interest implies a need for query strategies 
that cannot generally be managed
at the level of a single data provider.

There is a need for road-guides pointing to the 
most useful resources,
or to compilations or databases where information
can be found about these resources. 
Such guides have been made in the past, and are of very
practical help for the novice as well as the  trained user,
for example:
Andernach et al.\ \cite{AHM94},
Egret \& Heck \cite{waw},
Egret \& Albrecht \cite{amp2},
Heck \cite{eppa},
Grothkopf \cite{librarians},
Andernach \cite{ALDIA}.

In the present paper our aim is to address the questions
related to the collection, integration and interfacing
of the wealth of astronomical Internet resources,
and also to describe some strategies that have to be developed
for building cooperative tools which will be essential in the
research environment of the decade to come.

%
\section{Compilations of astronomical Internet resources}

At a first level, the user looking for new sources of
information can consult compilations
of existing resources.  Examples of such databases,
or yellow-page services are given in this section.

\subsection{The StarPages}

\emph{Star*s Family} is the generic name for a collection of 
directories, dictionaries and databases which has been 
described in details by Heck (\cite{starfam})
who has been building
up their contents for more than twenty-five years.   
These very exhaustive data sets are carefully updated and
validated, thus constituting a gold mine
for  professional, amateur astronomers, and more generally
all those who are curious of space-related activities,
and want to locate existing resources.

The Star*s Family of products can be 
queried on-line from the CDS Web site (Strasbourg, France)
under the generic name of 
StarPages\footnote{http://cdsweb.u-strasbg.fr/starpages.html}.
It includes the following databases: 
\begin{description}
\item[StarWorlds:]  a directory of astronomy, space sciences,
and related organizations (Heck et al.\ \cite{starworlds}); 
it includes URLs of Web sites
when available, as well as e-mail addresses;
unlike most of the services mentioned in
the present paper, it is not restricted to describing
on-line resources, but also lists directory entries
for organizations which do not provide any on-line
information.
\item[StarHeads:]   individual Web pages essentially of
astronomers and related space scientists
(Heck \cite{starheads}).
\item[StarBits:]  a very
comprehensive dictionary of abbreviations,
acronyms, contractions, and symbols used in astronomy and
space sciences (Heck \cite{starheads}). 
\end{description}
All three databases are associated with a query engine
based on character string searches. Filters prevent
extraction of too large subsets of the database.

\subsection{AstroWeb} 
\label{AstroWeb}

\emph{AstroWeb} (Jackson et al. \cite{astroweb}) is a 
collection of pointers to astronomically relevant 
information resources available on the Internet.
The browse mode of AstroWeb opens a window
on the efforts currently developed -- in some cases, unfortunately,
in a rather disorganized way -- for
making astronomically related, and hopefully pertinent,
information available on-line through the World Wide Web.

AstroWeb is maintained by a small consortium of individuals
located at CDS, STScI, MSSSO, NRAO, and Vilspa. The master
database is currently hosted at 
CDS\footnote{http://cdsweb.u-strasbg.fr/astroweb.html} 
(after having been
for a long time at STScI), and all the above-mentioned
places, as well as the Institute of Astronomy, Cambridge,
 host a mirror copy with
customized presentation of the same data.

Each URL is checked by a robot on a daily basis to ensure
aliveness of all referenced resources. The resource descriptions
are usually submitted by the person or organization responsible
for the resource, but are checked and eventually
modified by one of the consortium  members. 
The search engine is a  {\sc wais} search index. The index is
constructed from the resource descriptions, and also includes 
all the words contained in the referenced home page.  This
latter feature is quite powerful for bringing new names of
projects, topics, research groups, very quickly to the index.

Table \ref{astroweb} lists the resources present in
the AstroWeb database in December 1999.

\begin{table}
\caption{Resources listed in the AstroWeb database (December 1999).
    The number of resources (Web sites) in each category is given
  between parentheses.
    A number of resources appear in more than one category.}
\footnotesize
\begin{tabular}{r p{6cm}}
\hline 
Organizations  & Astronomy Departments (508) \\
                 & Professional and Amateur Organizations (159) \\
                 & Space Agencies and Organizations (46) \\
  \\
       Observing & Observatories and Telescopes (328) \\
      resources  & Telescope Observing Schedules (25) \\
                 & Meteorological Information (10) \\
                 & Astronomical Survey Projects (65)  \\
 \\ 
  Data resources & Data and Archive Centers (145) \\
                 & Astronomy Information Systems (39) \\
 \\
      Abstracts, & Bibliographical Services (29) \\
   Publications, & Astronomical Journals and Publications (90) \\
      Libraries  & Astronomy \& astrophysics preprints (58) \\
                 & Abstracts of Astronomical Publications (29) \\
                 & Conference Proceedings (45) \\
                 & Astronomy-related Libraries (48) \\
                 & Other Library resources (11) \\
 \\
People-related & Personal Web pages (800)  \\
      Resources  & People (lists) (14) \\
                 & Jobs (37) \\
                 & Conferences and Meetings (45)  \\
                 & Newsgroups (31)  \\
                 & Mailing Lists (16)  \\
 \\
        Software & Astronomy software servers (129) \\
        Computer & Document Preparation Tools (9) \\
        Science  & Overviews \& technical notes for protocols (11) \\
                 & Computer Science-related Resources (33) \\
 \\
  Research areas & Radio Astronomy (109) \\
      Astronomy  & Optical Astronomy (178) \\
  Space Physics  & High-Energy Astronomy (77) \\
                 & Space Astronomy (175) \\
                 & Solar Astronomy (77) \\
                 & Planetary Astronomy (64) \\
                 & History of Astronomy (21) \\
                 & Earth, Ocean, Atmosphere, Space Sciences (41) \\
                 & Physics-related Resources (91) \\
 \\
     Educational & Professional and Amateur Organizations (159) \\
      resources  & Educational resources (240) \\
                 & Astronomy Pictures (105) \\
 \\
Miscellaneous  & Primary Lists of Astronomy Resources (10) \\
                 & Other lists of astronomy resources (78) \\
                 & Miscellaneous Resources (137) \\
\hline
\end{tabular}
\label{astroweb}
\end{table}

\section{Current status of on-line astronomy resources}

Following the classification scheme adopted by AstroWeb,
we will outline in this section the current status of
the main categories of on-line astronomy resources,
pointing to meta-resources (i.e. organized lists of
resources) when they are available.

\subsection{Organizations}

Most of the active astronomical organizations (institutes,
astronomy departments, etc.) now have
home pages on the Internet. 
StarWorlds\footnote{http://cdsweb.u-strasbg.fr/starworlds.html} 
is currently the most comprehensive searchable directory of 
such resources ; it can be queried by names, keywords,
or character strings. 
For browsing lists sorted by alphabetical
order or by country, see AstroWeb (Section~\ref{AstroWeb}). 
National or international organizations also maintain
useful lists.

\subsection{Observational projects and missions}

It is now difficult to envisage an observational project
without a web site. As they are more dynamic and often
involve multiple organizations or institutions,
the best way to find
them may be to use one of the powerful commercial search
engines  that routinely index millions of web pages on the
Internet.

The indexing system of AstroWeb may also be 
helpful, especially when it is important to
limit the investigation domain to astronomy,
or to keep track of new emerging projects.

\subsection{Data and information systems}

Astronomy data and information centers are becoming increasingly 
interconnected, with both explicit links to other relevant resources and 
automatic cross-links that may be invoked transparently
to the end-user.
Section~\ref{isaia} describes current efforts to provide
interoperability within astrophysics (\emph{Astrobrowse}) and
across the space sciences (\emph{ISAIA}).

\subsection{Bibliographic resources}

Here also a virtual network is being organized,
as exemplified by the 
\emph{Urania}\footnote{http://www.aas.org/Urania/} 
initiative, or by the coordinated efforts to
create links between ADS and other services
(Kurtz et al.\ \cite{ADS}).
Note that many of the bibliographical resources are
electronic journals for which a subscription may be
required. 

\subsection{People-related resources}

Some databases 
(RGO E-mail directory\footnote{http://star-www.rl.ac.uk/astrolist/astrosearch.html},
StarHeads\footnote{http://cdsweb.u-strasbg.fr/starheads.hml})
follow the development
of electronic mail addresses and personal Web pages.
Directories from national or international societies
(e.g., AAS, EAS, IAU) are also
generally very carefully kept up to date.

The database of meetings and conferences maintained
by CFHT\footnote{http://cadcwww.dao.nrc.ca/meetings/meetings.html} is very
complete and well organized. Astronomical societies also
maintain their own lists.

\subsection{Astronomical software}

The Astronomical Software and Documentation Service 
(ASDS\footnote{http://asds.stsci.edu/}) 
is a network service that allows users to locate existing
astronomical software, associated technical documentation, and
information about telescopes and astronomical instrumentation
(Payne et al.\ \cite{ASDS96}).
ASDS originated as a service devoted entirely to 
astronomical software packages and their associated 
on-line documentation and was originally called the
Astronomical Software Directory Service.  Much code is rewritten these days, 
not because anyone has found a fundamentally better way 
to solve the problem, but because developers simply don't know 
who has already done it, whether the code runs on the 
system they have available, or where to get it if it does. 
That is the problem that ASDS was intended to solve. 

In 1998 the scope of ASDS was expanded to include 
astronomical observing sites and their associated 
telescope and instrument manuals, taken from a listing 
maintained at CFHT. The service was renamed at this point. 

\subsection{Educational resources}

Education and public outreach have always been a strong
concern in astronomy, but the importance of this
activity is growing at a higher rate, with the advent
of the World Wide Web.

It is difficult to give general rules for such a
wide field, going far beyond the limits of
astronomical institutions.  
Let us just say that we expect to see in the
future an increasing r\^ole of educational
institutions (planetariums, or outreach departments
of big societies or institutions), for conveying 
general astronomy knowledge, or news about recent 
discoveries, to the general public.

The yellow-page services mentioned above do keep
lists of the most important education services.

\section{Towards a global index of astronomical resources}

In the following we will focus on Internet resources that
actually provide data, of any kind, as opposed to those
describing or documenting an institution or a research
project, without giving access to  any data set or archive.

One main trend is certainly the increase of interconnections between
distributed on-line services, the `Weaving of the Astronomy Web' (which
was the title of a Conference organized in Strasbourg 
by Egret \& Heck \cite{waw}).

More generally, with the development of the Internet, and of a large
number of on-line services giving access to data or information, it is
clear that tools giving 
coordinated access to distributed services are needed. This
is, for instance, the concern expressed by NASA through the Astrobrowse
project (Heikkila et al.\ \cite{astrobrowse-2}). 

In this section we will first describe a tool for managing
a ``metadata'' dictionary of astronomy information services
(GLU); then we will show how the existence of such a
metadatabase can be used for building efficient search
and discovery tools.

\subsection{The CDS GLU}

The CDS (Centre de Donn\'ees astronomiques de Strasbourg) has 
recently developed a tool for managing remote links in a context of 
distributed heterogeneous services 
(GLU\footnote{http://simbad.u-strasbg.fr/glu/glu.htx}, 
G\'en\'erateur de Liens 
Uniformes, i.e. Uniform Link Generator; 
Fernique et al.
\cite{glu}). 

First developed for
ensuring efficient interoperability  of the several services existing at CDS 
({\sc VizieR},  {\sc Simbad}, {\sc Aladin}, bibliography, etc.; see
Genova et al.\  \cite{CDS}), 
this tool has also been 
designed for maintaining addresses (URLs) of 
distributed services 
(ADS, NED, etc.).

A key element of the system is the ``GLU dictionary'' maintained by 
the data providers contributing to the system, and distributed to all 
sites of a given domain. This dictionary contains knowledge about the 
participating services (URLs, syntax and semantics of input fields, 
descriptions, etc.), so that it is possible to generate automatically a 
correct query for submission to a remote database. 

The service provider (data center, archive manager, or  
webmaster of an astronomical institute) can use GLU for 
coding a query, taking benefit of the easy update of the system: 
knowing which service to call, and which answer to expect from 
this service, the programmer does not have to worry about the 
precise address of the remote service at a given time, nor of the 
detailed syntax of the query (expected format of the equatorial 
coordinates, etc.).

\subsection{New search and discovery tools}

The example of GLU demonstrates the usefulness of storing into
a database the knowledge about information
services (their address,  purpose, domain of coverage,
query syntax, etc.). In a second step, such a
database can be queried when the challenge is to provide
information about whom is providing what, 
for a given object, region of the sky, or domain of interest.

Several projects are working toward providing general solutions.

\subsubsection{Astrobrowse}

\emph{Astrobrowse} is a project that began within the United States
astrophysics community, primarily within NASA data centers, for developing
a user agent which significantly streamlines 
the process of locating astronomical data on the web. 
Several prototype implementations are already 
available\footnote{http://heasarc.gsfc.nasa.gov/ab/}.
With any of these prototypes, a user can already 
query thousands
of resources  without having to deal with out-of-date URLs, 
or spend time figuring out how to use each resource's
unique input formats. 
Given a user's selection of 
web-based astronomical databases and an 
object name or coordinates, Astrobrowse will 
send queries to all databases identified as containing potentially
relevant data.  It provides links to these resources and allows the
user to browse results from each query.  Astrobrowse does not recognize,
however, when a query yields a null result, nor does it integrate
query results into a common format to enable intercomparison.

\subsubsection{AstroGLU}

Consider the following scenario: we have a data
item $I$  (for example an author's name, the position or name
of an astronomical  object, a bibliographical reference, etc.),
and we would like to know  more about it, but we do not know a
priori which service $S$ to contact, and  what are the
different data types $D$ which can be  requested.
This scenario is typical of a scientist
exploring new domains as part of a research procedure.

The GLU dictionary can actually be used for helping to solve this 
question: the dictionary can be considered as a reference 
directory, storing the knowledge about all services accepting data item $I$ as 
input, for retrieving data $D_1$ or $D_2$.  For example, we can 
easily obtain from such a dictionary the list of all services accepting 
an author's name as input; information which can be accessed, in 
return, may be an abstract (service ADS), a preprint (LANL/astro-
ph), the author's address (RGO e-mail directory) or personal
Web  page (StarHeads), etc.

Based on such a system, it becomes possible to create automatically 
a simple interface guiding the user towards any of the services 
described in the dictionary.

This idea has been developed as a prototype tool, under the name of 
AstroGLU\footnote{http://simbad.u-strasbg.fr/glu/cgi-bin/astroglu.pl}
(Egret et al.\ \cite{astroglu}).
The aim of this tool is to help the users find their way among 
several dozens (for the moment) of possible actions or services.
A number of compromises have to be taken between providing the 
user with the full information (which would be too abundant and 
thus unusable), and preparing digest lists (which implies hiding some
amount of auxiliary information and making somewhat subjective
selections).

A resulting issue is the fact that the system puts on the same line 
services which have very different quantitative or qualitative 
characteristics.  {AstroGLU} has no
efficient ways yet to provide the user with a hierarchy of
services, as a gastronomic guide would do for restaurants.
This might come to be a necessity in the future, as more and more 
services become (and remain) available.

%
\section{Towards an integration of distributed data and information
  services}
\label{isaia}

 To go further, one needs
to be able to integrate the result of queries 
provided by heterogeneous services.
This is the aim of the ISAIA (Integrated System for Archival 
Information Access)
project\footnote{http://heasarc.gsfc.nasa.gov/isaia/}
(Hanisch \cite{isaia1}, \cite{isaia2}).

The key objective of the project is to
develop an interdisciplinary data location and integration 
service for space sciences. Building upon existing data
services and communications protocols, this service will 
allow users to transparently query a large variety of distributed
heterogeneous Web-based resources (catalogs, data, computational 
resources, bibliographic references, etc.)
from a single interface. The service will collect responses 
from various resources and integrate them in a seamless
fashion for display and manipulation by the user. 

Because the scope of ISAIA is intended to span the space 
sciences -- astrophysics, planetary science, solar physics, and
space physics -- it is necessary to find a way to standardize the
descriptions of data attributes that are needed in order to formulate
queries.  The ISAIA approach is based on the concept of \emph{profiles}.
Profiles map generic concepts and terms onto mission or dataset specific
attributes.  Users may make general queries across multiple disciplines
by using the generic terms of the highest level profile, or make more
specific queries within subdisciplines using terms from more detailed
subprofiles.

The profiles play three critical and interconnected roles:
\begin{enumerate}
\item They identify appropriate resources (catalogs, mission datasets,
bibliographic databases):  the \emph{resource profile}
\item They enable generic queries to be mapped unambiguously onto 
resource-specific queries:  the \emph{query profile}
\item They enable query responses to be tagged by content type and 
integrated into a common presentation format:  the \emph{response
profile}
\end{enumerate}
The resource, query, and response profiles are all aspects of a common
database of resource attributes.  Current plans call for these profiles to 
be expressed using XML (eXtensible Markup Language, an emerging standard
which allows embedding of logical markup tags within a document) and to be 
maintained as a distributed database using the CDS GLU facility.

The profile concept is critical to a distributed data service where one
cannot expect data providers to modify their internal systems or services
to accommodate some externally imposed standard.  The profiles act as a
thin, lightweight interface between the distributed service and the
existing specific services.  Ideally the service-specific profile
implementations are maintained in a fully distributed fashion, with
each data or service provider running a GLU daemon in which that site's
services are fully described and updated as necessary.  Static services 
or services with insufficient staff resources to maintain a local GLU
implementation can still be included, however, as long as their profiles
are included elsewhere in the distributed resource database.  The profile
concept is not unique to space science, but would apply equally well to 
any distributed data service in which a common user interface is desired
to locate information in related yet traditionally separate disciplines.

%
\section{Information clustering  and advanced user interfaces}

A major challenge in current information systems research is to find
efficient ways for users to be able to visualize the contents and understand
the correlations within large databases.  The technologies being developed
are likely to be applicable to astronomical information systems.  For
example,
information retrieval by means of ``semantic road maps'' was first 
detailed in Doyle (\cite{doyle}), using a powerful
spatial metaphor which 
lends itself quite well to modern distributed computing 
environments such as 
the Web. 
The Kohonen self-organizing feature map (SOM;
Kohonen \cite{origk}) method 
is an effective means towards this end of a
visual information retrieval user interface.  

\subsection{Interfacing datasets with a Self-organizing Map}

The Kohonen map is, at heart, $k$-means clustering with the additional 
constraint that cluster centers be located on a regular grid (or some 
other topographic structure) and furthermore their location on the grid 
be monotonically related to pairwise proximity (Murtagh \& 
Hern\'andez-Pajares \cite{mhp}).

A regular grid is quite convenient for
an output representation space,
as it maps conveniently onto a visual user interface.
In a web context, it can easily be made interactive and responsive.

\begin{figure}
\resizebox{\hsize}{!}{\includegraphics{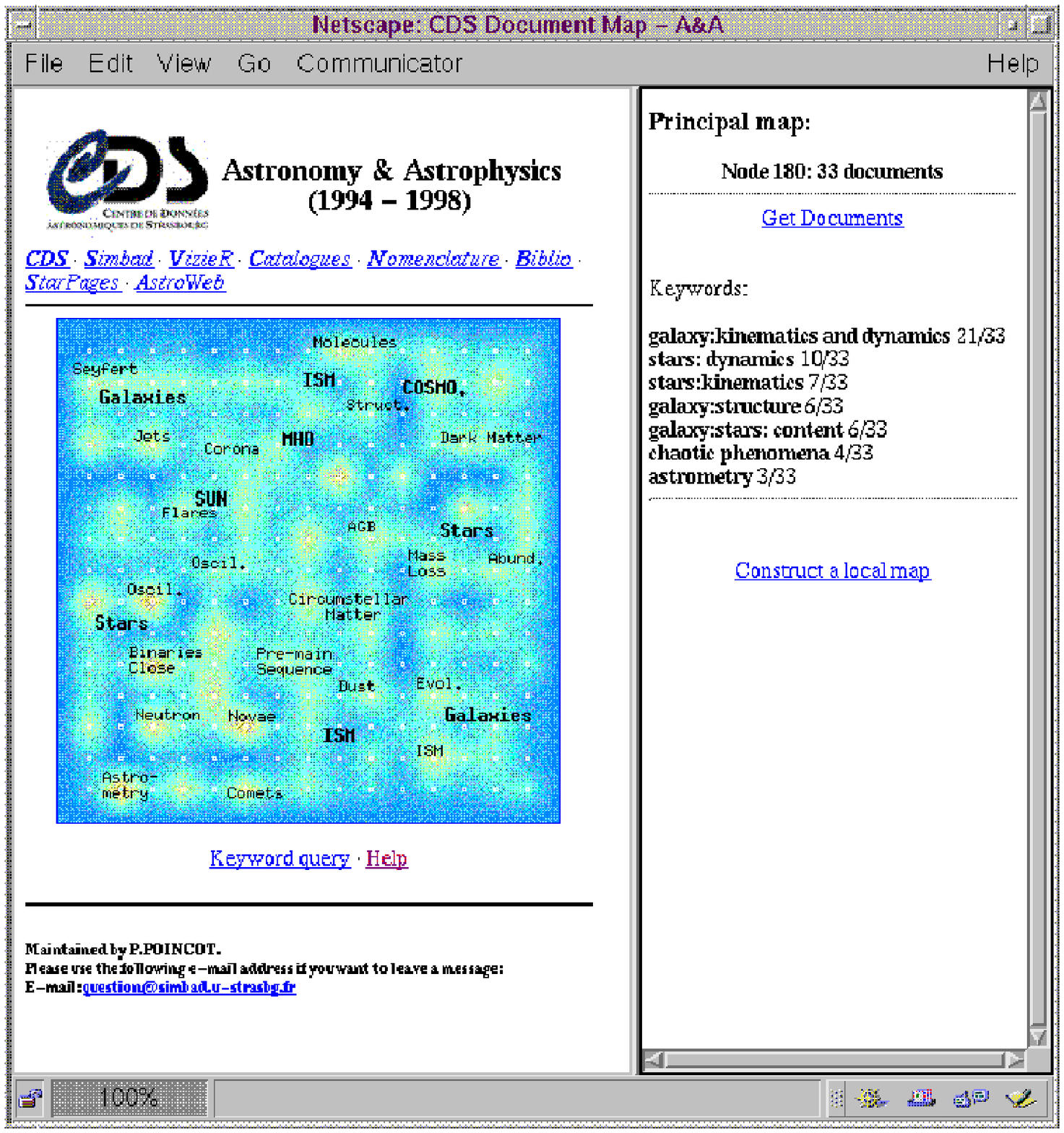}}
\caption{Visual interactive user interface to a set
of articles from the journal 
{\em Astronomy and Astrophysics}.  Original in color.}  
\label{figmap}
\end{figure}

Fig.\ \ref{figmap} shows an example of such
a visual and interactive user interface  map,
in the context of a set of journal articles described
by their keywords.
Color is related to density of document clusters located at 
regularly spaced nodes
of the map, and some of these nodes/clusters are annotated.  
The map is installed on the Web as a clickable image map, 
with CGI programs accessing lists of documents
and -- through further links -- in many
cases, the full documents. 
In the example shown, the  user has queried a node and
results are seen in the right-hand panel. 
Such maps are maintained for (currently) 
12,000 articles from the {\em Astrophysical Journal}, 7000 from 
{\em Astronomy and Astrophysics}, over 2000 astronomical catalogs, and 
other data holdings.    More information on the design of this
visual interface and 
user assessment can be found in Poin\c{c}ot et al.\ (\cite{poin1,poin2}).

\subsection{Hyperlink clustering}

\begin{figure}
\resizebox{\hsize}{!}{\includegraphics{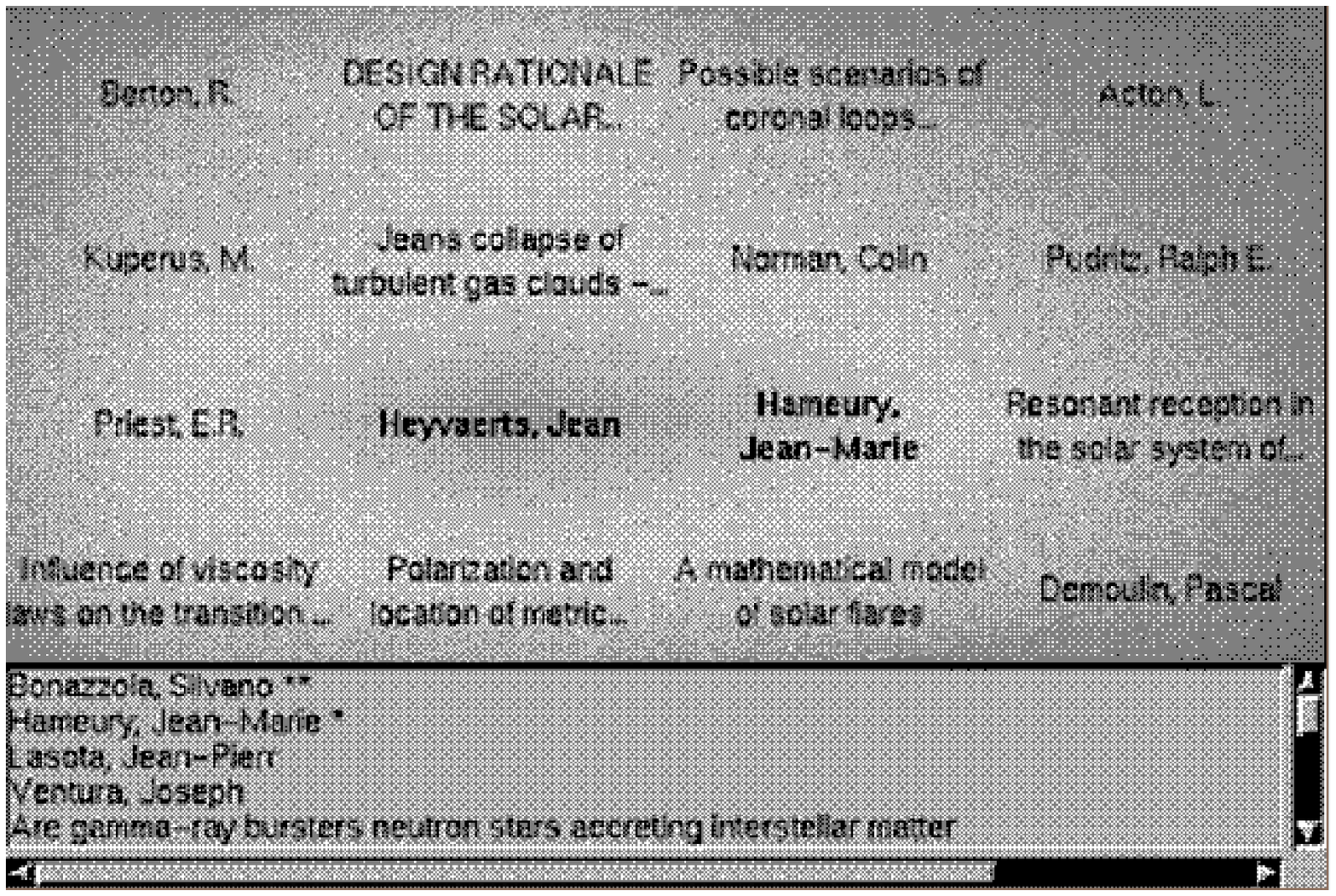}}
\caption{Visual interactive user interfaces, based on graph edges. Vertices
are author names, article titles and (not shown here) astronomical object
names.   Map for astronomer Jean Heyvaerts.  Original in color.}  
\label{heyvaerts}
\end{figure}

Guillaume \& Murtagh (\cite{guill}) have recently
developed a Java-based visualization
tool for hyperlink-based data, in XML,
consisting of astronomers, astronomical
object names, article titles, and possibly other objects
(images, tables, etc.).  
Through weighting, the various types of links could
be prioritized.  An iterative refinement algorithm was developed to map the
nodes (objects) to a regular grid of cells, which, as for the Kohonen SOM 
map, are clickable and provide access to the data represented by the 
cluster.  
Fig.\ \ref{heyvaerts} shows an example for an astronomer 
(Prof.\ Jean Heyvaerts, Strasbourg Astronomical Observatory).

These new cluster-based visual user interfaces are not computationally 
demanding.  In general they cannot be created in real time, but 
they are scalable in the sense that many tens of thousands of documents or
other objects can be easily handled.  
Document management (see e.g.\ Cartia\footnote{http://www.cartia.com/}) 
is less the motivation as is 
instead the interactive user interface.  

Further information on these visual user interfaces can be found in 
Guillaume (\cite{guill2}) and Poin\c{c}ot (\cite{poin3}).  

\subsection{Future developments for advanced interfaces}

Two directions of development are planned in the near future.  
Firstly,
visual user interfaces need to be coupled together.  A comprehensive 
``master'' map is one possibility, but this has the disadvantage of 
centralized control and/or configuration control.  
Another possibility is to develop a protocol 
such that a map can refer a user to other maps in appropriate circumstances.
Such a protocol was developed a number of years ago in a system called
Ingrid\footnote{http://www.ingrid.org/} 
developed by P.\ Francis at NTT Software 
Labs in Tokyo (see Guillaume  \cite{guill2}).  However this work has been 
reoriented since then. 

Modern middleware tools may offer the 
following solution.
This is to define an information 
sharing bus, which will connect distributed information maps.  It
will be interesting to
look at the advantages of CORBA (Common Object Request Broker Architecture)
or, more likely, EJB (Enterprise Java Beans), for ensuring this 
interoperability infrastructure (Lunney \& McCaughey \cite{corba}). 

A second development path is to note the clustering which is at the core
of these visual user interfaces and to ask whether this can be further
enhanced to facilitate construction of query and response agents.  It is 
clear to anyone who uses Internet search engines such as AltaVista, 
Lycos, etc.\ that clustering of results
is very desirable.  A good example of such clustering of search results in
practice is the Ask Jeeves search engine\footnote{http://www.askjeeves.com/}.
The query interface, additionally, is a natural language one, another
plus.

\section{Conclusion}

The on-line ``Virtual Observatory'' is currently
under construction with on-line archives and services 
potentially giving access to a huge quantity
of scientific information: its services will allow astronomers to
select the information of interest for their research, and to access
original data, observatory archives and results published in
journals. Search and discovery tools currently in 
development will be of vital importance to make all the
observational data and information
available to the widest scientific community.

\begin{acknowledgements}

CDS acknowledges the support of INSU-CNRS, the Centre National
d'Etudes Spatiales (CNES), and Universit\'e Louis Pasteur.
\end{acknowledgements}



\begin{thebibliography}{}

\bibitem[1994]{AHM94}
Andernach, H., Hanisch, R.J., Murtagh, F., 1994, PASP 106, 1190

\bibitem[1999]{ALDIA}
Andernach, H., 1999, in 
\emph{Astrophysics with Large Databases in the Internet Age},
Proc. IXth Canary Islands Winter School on Astrophysics,
M. Kidger, I. P\'erez-Fournon, \& F. S\'anchez (Eds.),
Cambridge University Press, p. 1



\bibitem[1961]{doyle}
Doyle, L.B., 1961, 
{\em Journal of the ACM}, 8, 553

\bibitem[1994]{jungle}
 Egret, D., 1994, in {\em Astronomical Data
Analysis Software and Systems III}, ASP Conf. Ser. 61, p. 14 

\bibitem[1995]{amp2}
Egret, D., Albrecht, M. (Eds.), 1995,
{\em Information \& On-line Data in Astronomy},
Kluwer Academic Publ., Dordrecht

\bibitem[1995]{waw}
Egret, D., Heck, A. (Eds.), 1995,
\emph{Weaving the Astronomy Web}, Vistas in Astron. 39, 1--127


\bibitem[1998]{astroglu} 
Egret, D., Fernique, P., Genova, F., 1998, 
in {\em Astronomical Data
Analysis Software and Systems VII}, ASP Conf. Ser. 145, p. 416
(AstroGlu)

\bibitem[1998]{glu} 
Fernique, P., Ochsenbein, F., Wenger, M., 1998, 
in {\em Astronomical Data
Analysis Software and Systems VII}, ASP Conf. Ser. 145, p. 466
(GLU)

\bibitem[2000]{CDS}  Genova, F., Egret, D., Bienaym\'e, O.,
et al., 2000, A\&AS, \emph{in press}
(CDS)

\bibitem[1995]{librarians}
Grothkopf, U., 1995,
Vistas Astron. 38, 401

\bibitem[2000]{guill}
 Guillaume, D., Murtagh, F., 2000, ``Clustering of XML documents,'' 
{\em Computer Physics Communications}, in press

\bibitem[2000]{guill2}
Guillaume, D., 2000, 
PhD thesis, Universit\'e Louis Pasteur, Strasbourg

\bibitem[2000a]{isaia1}  Hanisch, R.J., 2000a, {\em Computer
Physics Communications}, in press

\bibitem[2000b]{isaia2}
Hanisch, R.J., 2000b, in {\it ADASS IX}, ASP Conf. Ser. in press

\bibitem[1995a]{starfam}
  Heck, A., 1995a, 
in {\em Information \& On-line Data in Astronomy}, D. Egret 
and M. A. Albrecht, Eds., p. 195
(Star*s Family)

\bibitem[1995b]{starheads}
  Heck, A., 1995b, A\&AS 109, 265
(StarHeads)

\bibitem[1997]{eppa}
  Heck, A., 1997, \emph{Electronic Publishing for Physics and Astronomy},
  Astrophys. Space Science 247, Kluwer, Dordrecht

\bibitem[1994]{starworlds}
  Heck, A., Egret, D., Ochsenbein, F., 1994, A\&AS 108, 447
(StarWorlds -- StarBits)

\bibitem[1999]{astrobrowse-2}
Heikkila, C.W., McGlynn, T.A., White, N.E., 1999,
in {\em Astronomical Data
Analysis Software and Systems VIII}, ASP Conf. Ser. 172, p. 221
(Astrobrowse)

\bibitem[1994]{astroweb} 
  Jackson, R., Wells, D., Adorf, H.M., et al., 1994, A\&AS 108, 235
(AstroWeb)

\bibitem[1982]{origk}
Kohonen, T., 1982, Biological Cybernetics 43, 59

\bibitem[2000]{ADS}  Kurtz, M., Eichhorn, G., Accomazzi, A.,
et al., 2000, A\&AS, in press
(ADS)

\bibitem[2000]{corba} Lunney, T.F., McCaughey, A.J., 1999,
   Computer Physics Communications, in press


\bibitem[1995]{mhp} Murtagh, F., Hern\'andez-Pajares, M., 1995,
Journal of Classification, 12, 165

\bibitem[1996]{ASDS96} Payne, H. E., Hanisch, R. J., Warnock, A., 1996,
in {\em Astronomical Data Analysis Software and Systems V},
ASP Conf. Ser. 101, 577

\bibitem[1998]{poin1}  
Poin\c{c}ot, Ph., Lesteven, S., Murtagh, F., 1998, A\&AS 130, 183

\bibitem[2000]{poin2} Poin\c{c}ot, Ph., Lesteven, S., Murtagh, F.,
2000,  
Journal of the American Society for
Information Science, {\em submitted}

\bibitem[1999]{poin3}
Poin\c{c}ot, Ph., 1999, PhD thesis, Universit\'e Louis Pasteur,
  Strasbourg


\end{thebibliography}
\end{document}